\documentclass[12pt]{nature}

\usepackage{times}
\usepackage[utf8]{inputenc}
\usepackage{gensymb}
\usepackage{graphicx}
\usepackage{lineno}
\usepackage{hyperref} 
\usepackage{multirow}
\usepackage{braket}
\usepackage{amsmath}
\usepackage[T1]{fontenc}
\usepackage{bm}
\usepackage{xcolor}
\usepackage{textcomp}
\usepackage[normalem]{ulem}
\usepackage{times}
\title{Axial Higgs Mode Detected by Quantum Pathway Interference in RTe$_{3}$}
\author{Yiping Wang$^{1}$, Ioannis Petrides$^{2}$, Grant McNamara$^{1}$, Md Mofazzel Hosen$^{1}$, Shiming Lei$^{3}$, Yueh-Chun Wu$^{4}$, James L. Hart$^{5}$,  Hongyan Lv$^{6}$,  Jun Yan$^{4}$, Di Xiao$^{7,8}$,  Judy J. Cha$^{5}$, Prineha Narang$^{2}$, Leslie M. Schoop$^{3}$, Kenneth S. Burch$^{1*}$}
\begin{document} 

\maketitle
\begin{affiliations}
\item{Department of Physics, Boston College, Chestnut Hill, MA 02467, USA}
\item{John A. Paulson School of Engineering and Applied Sciences, Harvard University, Cambridge, MA 02138, USA}
\item{Department of Chemistry, Princeton University, Princeton, NJ 08544, USA}
\item{Department of Physics,University of Massachusetts Amherst, Amherst, MA 01003 USA}
\item{Department of Mechanical Engineering and Materials Science, Yale University, New Haven, CT 06511, USA}
\item{Key Laboratory of Materials Physics, Institute of Solid State Physics, HFIPS, Chinese Academy of Sciences, Hefei 230031, China}
\item{Department of Materials Science and Engineering, University of Washington, seattle, WA 98105, USA}
\item{Department of Physics, University of Washington, seattle, WA 98105, USA}\\
{$^*$ To whom correspondence should be addressed; E-mail:  ks.burch@bc.edu}
\end{affiliations}

% Double-space the manuscript.
\baselineskip24pt

\newcommand*\rte{{RTe$_3$}}
\newcommand*\gdte{{GdTe$_3$}}
\newcommand*\late{{LaTe$_3$}}
\begin{abstract}
The observation of the Higgs boson solidified the standard model of particle physics. However, explanations of anomalies (e.g. dark matter) rely on further symmetry breaking calling for an undiscovered axial Higgs mode\cite{Franzosi2016}. In condensed matter the Higgs was seen in magnetic, superconducting and charge density wave(CDW) systems\cite{Shimano,Varma2015}. Uncovering a low energy mode's vector properties is challenging, requiring going beyond typical spectroscopic or scattering techniques. Here, we discover an axial Higgs mode in the CDW system \rte\ using the interference of quantum pathways. In \rte\ (R=La,Gd), the electronic ordering  couples bands of equal or different angular momenta\cite{Leslie_jacs.0c01227,Brouet_PhysRevB.77.235104,Leieaay6407}. As such, the Raman scattering tensor associated to the Higgs mode contains both symmetric and antisymmetric components, which can be excited via two distinct, but degenerate pathways. This leads to constructive or destructive interference of these pathways, depending on the choice of the incident and Raman scattered light polarization.  The qualitative behaviour of the Raman spectra is well-captured by an appropriate tight-binding model including an axial Higgs mode. The elucidation of the antisymmetric component provides direct evidence that the Higgs mode contains an axial vector representation (i.e. a psuedo-angular momentum) and hints the CDW in \rte\ is unconventional. Thus we provide a means for measuring collective modes quantum properties without resorting to extreme experimental conditions. 

%When continuous symmetry of a system is spontaneous broken, two types of collective modes, Nambu-Goldstein and Higgs(amplitude) mode, would emerge. The measurement of collective mode quantum properties typically requires extreme experimental conditions: low temperatures, ultrafast lasers, high magnetic and/or electric fields. Dispite the Higgs mode has been observed in charge density wave(CDW) systems for decades, it is always assumed to be a scalar as CDW are typically assumed to have s-wave condensation. 

%Specifically, the axial Higgs mode is observed for incident polarization along a crystal axis with perpendicular scattered polarization but disappears upon just swapping the polarizations, while the intensity of all phonons modes remains unchanged. 

\end{abstract}
Emergent order states bring about new modes whose properties are directly linked to the associated change in topology or symmetry breaking. A well studied example is the breaking of translation symmetry in CDW systems resulting in Nambu-Goldstone (phase) and Higgs (amplitude) modes.  In a superconductor the inclusion of electromagnetism results in a gapping of the phason or giving mass to the W and Z bosons of the electroweak theory. However this Anderson-Higgs mechanism leaves the amplitude mode largely unchanged, and thus by convention it is often referred to as the Higgs boson.\cite{Shimano,Podolsky2011VisHigg,Varma2015} Despite its close resemblance to superconductivity and first prediction in 1955, all CDW to date have revealed s-wave condensation with a scalar Higgs mode. The challenge in detecting unconventional order is the requirement to probe the vector nature of the order parameter or collective excitations. One example is the attempt to extend the standard model by enlarging the symmetry breaking. This results in additional particles, including a spin-1 Higgs boson, and dark matter candidates.\cite{Franzosi2016} Thus the detection of a Higgs mode with finite angular momenta (i.e. an axial Higgs) heralds the discovery of a heretofore undiscovered symmetry breaking and novel phase of matter. 

Pathway interference is an elegant tool to meet this challenge by exploiting wave-particle duality to uncover the hidden quantum properties of excitations\cite{zeilinger1988single}. In condensed matter quantum pathway interference revealed the topological properties of band structures\cite{yuanboZhang2005,Qu821}, unconventional superconducting order\cite{Ryu2020,Cleuziou2006,Giazotto2010}, and the nontrivial statistics of collective excitations\cite{Mittal2021,Wall2011,Barik666}. Despite the elegance of such experiments, they have not been directly applied to the Higgs mode to uncover its axial vector nature. In part, this results from the challenge of performing quantum interference in condensed matter settings, which typically require extreme experimental conditions: low temperatures, ultrafast lasers, high magnetic and/or electric fields. The need for such conditions could be overcome by CDW systems, with well defined Higgs and phase modes readily observed with optical techniques at large energy scales\cite{Popescu2010,Chang2012_nphys,PRB_Raman_LM,Kogar2020,Brouet_PhysRevB.77.235104,Yusupov_PhysRevLett.101.246402_Pump_probe,Liu.2013}. Furthermore, the CDW can be tuned by pressure, exfoliation, or ultrafast lasers\cite{zocco_PhysRevB.91.205114,Xi2015_Nbse,Yoshikawa2021,Vaskivskyi2015} while offering next generation nano oscillator, logic and memory devices\cite{Balandin2021}. 

Here we study quantum pathway interference of the axial Higgs mode to reveal the unconventional CDW phase in \rte. This is achieved at room temperature with Raman scattering as it can measure the energy, symmetry, and excitation pathways of fundamental modes\cite{Klein_PhysRevB.25.7192,Wang2020,Devereaux_RevModPhys.79.175,ralston1970resonant}. We build upon previous inelastic light scattering experiments in non-interacting systems revealed the chiral nature of phonons, molecular crystal field excitations, semiconductor interband transitions, and changes in quantum pathway or coherence upon gating\cite{cardona_1975,Koningstein.1968,FengWang_graphene,xiaodongxu2015}. For these previous single particle experiments, the intermediate states (i.e. quantum pathways) are chosen by the combination of excitation wavelength, momentum conservation, the polarization of the incoming and outgoing light via selection rules\cite{friedman1975interference,chen1990interference}.

We focused on the rare earth CDW system \rte\ (R = Gd,La) exploiting its high transition temperatures ($T^{Gd}_{CDW}=380$~K, $T^{La}_{CDW}> 600$ K), unidirectional CDW, and multiple nesting conditions, enabling quantum pathway selection. The \rte\ crystal structure consists of double layers of van der Waals bonded square-planar Te sheets separated by RTe slabs (Fig. 1a), crystallized in an orthorhombic structure with the space group Bmmb. \rte\ is nearly tetragonal ($a - b \approx 0.01$\AA)\cite{Malliakasja0641608,Leieaay6407} with an incommensurate CDW propagating along the b axis, which is confirmed via transmission electron microscopy (TEM) in the supplemental (Fig. S1). 
 The Bloch bands near the Fermi energy are mainly composed of the $P_{X}$ and $P_{Y}$ orbitals of the tellurium sheet (Fig. 1b) as the $P_{Z}$ orbital is much lower in energy. Since the \rte\ slab is less densely packed within the ab plane, the chosen unit cell (c.f., Fig. 1a) results in the Fermi surface shown in Fig.1c, where hybridization between the two orbitals happens only at isolated points due to next-nearest neighbour interaction~\cite{Eiter64}.
The incommensurate $q_{CDW}$ determined by various techniques is $(2/7)b^{*}$, where $b^{*}$ is the reciprocal lattice vector, consistent with nesting between the original $P_{X}$ derived band of the Te sheet and the $P_{X}$ band folded in due to the enlarged unit cell of the 3D  structure\cite{Leslie_jacs.0c01227,Brouet_PhysRevB.77.235104}. In addition, another nesting condition arises when accounting for the reciprocal lattice vector, where $b^{*}-q_{CDW}$ connects $P_X$($P_Y$) to $P_{Y}$($P_{X}$) derived bands. Due to the orthorhombicity, these nesting conditions are not satisfied along the a-axis, resulting in a node in the CDW gap,  suggesting an unconventional order. Similarly the requirement to change angular momenta when connecting the $P_{X}$ to $P_{Y}$ states with $b*-q_{CDW}$ suggests the Higgs mode has finite angular momenta. The interference between the two pathways associated with this mode, the possible axial nature of the Higgs, change in sign of the gap (i.e. p-wave order) and thus unconventional CDW order have yet to be explored theoretically or experimentally.

As seen in Fig. 2 a-b, the intermediate states have either the same or different angular momenta and thus follow different selection rules. The selection rules depend sensitively on polarization relative to the crystal axis, presence of single domains and clean surfaces. We achieved this with exfoliated \rte\ flakes produced in a glovebox and rapidly transferred via vacuum suitcase to our low T Raman system. This ensured the sample surface is free from oxide contamination, atomically flat, contained single CDW domains and enabled identification of the crystal axes using the sharp edges (confirmed via TEM -- see supplemental). 

With this in mind, we turn to the polarization dependent Raman spectra of the Higgs mode. Fig. 1d shows representative measurements using a 532~nm excitation of \gdte\ at 300K in both parallel and cross polarization configurations. Here ab(XY 0$^o$) refers to the incident (scattered) light polarized along the crystal a(b)-axis. Similarly, a'b'(XY 45$^o$) represents the crystal rotated by 45 degrees from the ab configuration, where a' = a+b, b' = a-b. The 5 meV broad peak is the CDW Higgs mode and all other sharp peaks are the phonons.

The Higgs mode partially overlaps with the phonons in the shaded energy region. As seen in (Fig.1d), both the phonons and Higgs mode are observed in parallel polarization along aa (XX 0 deg), bb (XX 90deg) or a'a'(XX 45deg) directions. As expected and typically observed, the phonon modes have the same intensity when the configuration is changed from a'b' (XY 45deg) to b'a'(XY 135deg). Indeed, the measured Raman intensities ($I$) for a given excitation are proportional to the square of the product of incident light polarization ($\hat{e}_i$), Raman tensor ($R$) and the scattered electric field ($\hat{e}_{f}$): $I = |\hat{e}_i\cdot R \cdot \hat{e}_f|^2$. However, as seen in the shaded region of (Fig.1d), the scattering intensity of the CDW Higgs mode behaves quite differently, it is strong in a'b' but dramatically reduced in b'a'. Other than the CDW Higgs mode, phonons coupling with CDW mode at 7.4 meV and 10.6 meV also showed the intensity difference at a'b' and b'a' polarization. As previously established in X-ray measurement, the Higgs mode is strongly mixed with these phonons\cite{PhysRevB.X_ray} and they disappear above CDW transition temperature\cite{Eiter64}. (Details in supplemental Figure .S6).

To understand the change in the Higgs mode intensity upon swapping the incident and scattered polarization, it is useful to consider the role of the CDW in the quantum pathways in \rte. From symmetry, the possible inelastic, $q\approx 0$ excitations must fall in one of the irreducible representations $\Gamma_{\text{Raman}} = 3 A_g + B_{1g}+ B_{2g}+ B_{3g} $ leading to a symmetric Raman tensor ($R_{ij} = R_{ji}$)\cite{powell2010symmetry}. As seen in Fig. 2 and 3, this produces a four-fold angular dependence of the phonon modes intensity. However, due to the periodicity of the CDW, there are two quantum pathways that involve different intermediate states separated by $|q_{CDW}|$ or $|b^* - q_{CDW}|$ (see Fig. 1c, 2a and b). In the first, an $X$($Y$) polarized incident photon excites the electron into an intermediate state $|P_X\rangle $($|P_Y\rangle $), which is scattered to the $|P_X\rangle $ ($|P_Y\rangle $) state by the Higgs mode with wave vector $q_{CDW}$. Subsequently, the electron recombines with a hole and emits an $X$($Y$) polarized photon. Such a process results in a symmetric response as it involves scattering between states with identical polarization. While the symmetric response could be a Raman tensor with the form of $A_{g}$, $B_{g}$ or a sum of the two,  in fitting the angular dependent Higgs mode susceptibility, we find that the off-diagonal symmetric terms ($B_{g}$ vertex) are nearly zero.

The second scattering pathway involves a Higgs mode connecting $|P_X\rangle $($|P_Y\rangle $) to states with different angular momenta $|P_Y\rangle$($|P_X\rangle $) via $|b*- q_{CDW}|$. The change in angular momenta of the states associated with this nesting vector suggests that the Higgs mode is unconventional, requiring an axial vector representation. Nonetheless, upon recombination, a $Y$ ($X$) polarized photon is produced. Noting that the excitation from the $|P_Z\rangle $ band to $|P_X\rangle$ or $|P_Y\rangle$ bands matches the visible excitation laser energy,\cite{Brouet_PhysRevB.77.235104} the Higgs mode is a resonant electronic response. This resonance combined with the angular momentum change induces a nonzero antisymmetric (i.e. $R_{ij}=-R_{ji}$) contribution to the Raman tensor~\cite{cardona_1975}, which by itself would produce a signal only in XY and not XX configurations. Ultimately it is the interference of this antisymmetric process with the symmetric, diagonal component that produces the two-fold response in the cross-polarized Raman. In the supplemental material we calculate the Raman susceptibility from resonant processes using a low energy model in the presence of a CDW gap and amplitude mode and find an asymmetric transition, depending on the chosen pathway. This asymmetry primarily comes from the points in the Fermi surface with nesting vector $b*-q_{CDW}$ and is enhanced by the orbital mixing due to next-nearest neighbour interaction~\cite{Eiter64}.
Fitting to the experimental data we arrive at a Raman tensor for the Higgs mode:
$$R_{CDW} = 
\begin{pmatrix}
0 & d & 0\\
-d & 0 & 0\\
0 & 0 & 0\\
\end{pmatrix}
+
\begin{pmatrix}
e & 0 & 0\\
0 & f & 0\\
0 & 0 & g\\
\end{pmatrix}
=
\begin{pmatrix}
e & d & 0\\
-d & f & 0\\
0 & 0 & g\\
\end{pmatrix}
$$
, where d, e, f, g are independent coefficients. The detailed calculations of the angular dependence of the Raman response are shown in the supplemental but we briefly describe the key results here. The $|P_X\rangle\rightarrow |P_X\rangle$ pathway gives the same response for a'b' and b'a' configurations: $I=|(e-f)|^{2}$. However, $|P_X\rangle\rightarrow |P_Y\rangle$ pathways give $I=|2d|^{2}$ under a'b' polarization and $I=|-2d|^{2}$ under b'a' polarization. As such if the two pathways did not interfere (i.e. we add their intensities) we would not observe any difference when swapping the incident and scattered polarization (i.e. four-fold response of Raman vertex diagonal term and off-diagonal symmetruc term - Fig. 2d ). However the indistinguishablity and thus pathway interference leads to the CDW intensity $I_{a'b'} = |(e-f)+2d|^2$ producing the constructive interference term and $I_{b'a'} = |(e-f)-2d)|^2$ the destructive interference term, and thus a purely two fold angular response in cross-polarized Raman (2-fold response of Raman vertex diagonal term and off-diagonal anti-symmetric term Fig. 2c).

To reveal the suggested quantum interference of pathways, we focus on the angular dependence of the Raman response (details in supplemental). The colormaps in Fig.3a and b present parallel and cross polarization data and the green lines are the selected spectra in Fig.1d. The shaded region indicates the response from the Higgs mode. Consistent with quantum interference, these maps clearly show the Higgs mode has a two fold response. To examine in detail the Higgs and phonon mode angular dependence, we plot the intensity versus crystal angle for representative phonons of A$_{g}$ and B$_{g}$ symmetry alongside the response from the Higgs mode (Fig.3c--e). In the supplemental material we derive the Raman tensor associated to each mode using a generic representation in the orthorombic crystal group where the parameters are optimized to best fit the experimental data. We find that both the A$_{g}$ and B$_{g}$ phonons follow the expected angular dependence with four fold modulation when the associated Raman tensor is purely symmetric, reflecting the fact that the structure is nearly tetragonal.

On the other hand, the Higgs mode reveals clear two fold modulated intensities in both parallel and cross polarization (Fig. 3e) that is well-described by the Raman tensor $R_{CDW}$, where both pathways are summed. To the best of our knowledge this is the first such observation in any Raman experiment. This result highlights the utility of the full angular dependence of the Raman in revealing and potentially controlling the quantum pathways in a CDW system. Indeed, by simply rotating the light polarization we vary the different pathways relative phase from 0 to $\pi$ (i.e. constructive and destructive interference).

We now discuss alternative origins of the anomalous Higgs response. One is intrinsic angular momentum from the Gd moments. Another could be the competing phase with a secondary CDW in \gdte. This is seen by tuning the rare earth, where a bidirectional CDW appears at low temperatures for rare earths smaller than Gd. The role of fluctuations of this bidirectional CDW are unclear. In our TEM measurements some \gdte\ flakes revealed very weak secondary CDW (see supplemental). 

Therefore to eliminate the complexities from Gd (magnetism and multidomains), we tested the response of \late, which possesses $T_{CDW}>$600~K and contains no magnetic moments. Nonetheless, since all the \rte\ have a similar electronic structure\cite{Brouet_PhysRevB.77.235104}, the quantum pathway interference should remain. As shown in Fig.4A, the angular dependence of the Higgs mode in \late\ also reveals a two fold symmetry in both parallel and cross polarization, while the B$_{g}$ phonon shows the expected four fold response (Fig. 4b). The \late\ result thus confirms the interference is not due to intrinsic moments or competing phases, but from the band structure and quantum pathway selection. To ensure the reproducibility and intrinsic nature of our results, we also tested another flake, exfoliated from a different \gdte\ crystal from another growth using a separate Raman setup with shorter wavelength excitation laser (488 nm). Since the 488 nm is still in resonance, it reveals the same angular dependence of the modulation of the intensity (Fig. 4c).

As a final check of the robustness of the quantum interference, we turned to the temperature dependence. Due to a large change in the Higgs mode energy, its interaction with a nearby phonon varies with temperature\cite{PRB_Raman_LM}. As shown in Fig. 4e and consistent with previous measurements, the Higgs mode softens from 10 meV to 3 meV when increasing temperature from 8~K to 300~K, along with a decrease in intensity. This is a typical temperature dependence for a Higgs mode, resulting from the potential landscape of the free energy being reduced as the CDW transition temperature is approached. In addition, the Higgs mode displays an avoided crossing with the 7.4 meV phonon consistent with it revealing the the same symmetry as the Higgs (see supplemental Fig. S6). This is due to the fact that this mode is folded to $q=0$ by the CDW and thus can undergo the same quantum pathway interference\cite{PRB_Raman_LM,APL_Gdte3,PhysRevB.X_ray}. Therefore, we choose three temperatures to test the quantum interference via the cross polarization modulated intensities: 8K, slightly below (140K) and well above the avoided crossing (300K). As seen in the polar plot in Fig. 4d, the Higgs mode has the exact same angular dependence at all temperatures. This demonstrates that the quantum interference is robust to the mixing of the Higgs mode with nearby phonons.

Our study provides the first detection of an axial Higgs mode, exploiting the quantum pathway interference in Raman scattering. The finite angular momentum of the Higgs provides compelling evidence that the charge density wave order in \rte\ is unconventional. Using a phenomenological description of the Fermi surface we elucidate the role of next-nearest neighbour interaction in the observed asymmetry of Raman transitions involving a change of angular momentum. The methodology employed, can be applied to search for new symmetry broken and topologically ordered states via their novel collective modes. Furthermore the straightforward application of tuning the interference with light polarization could enable manipulating the quantum properties of collective excitations towards new non-equilibrium states.

\section*{Methods}
\subsection{Crystal Growth}
High-quality \rte\ single crystals were grown in an excess of tellurium (Te) via a self flux technique. Te (metal basis >99.999\%, Sigma-Aldrich) was first purified to remove oxygen contaminations and then mixed with the rare-earth (>99.9\%, Sigma-Aldrich) in a ratio of 97:3. The mixture was sealed in an evacuated quartz ampoule and heated to 900 C over a period of 12 hours and then slowly cooled down to 550\degree C at a rate of 2\degree C/hour. The crystals were separated  from the flux via centrifugation at 550\degree C

\subsection{Sample Preparation and Vacuum Transfer}
\gdte\ and \late\ flakes were exfoliated and characterized using unpolarized Raman in an Argon glovebox. Then they were loaded into a $10^{-6}$ mbar vacuum suitcase and directly transferred into a low temperature cryostat\cite{Mason_RSI}.

\subsection{Angle Resolved Raman Spectroscopy}
The 532nm Raman experiments were performed with a custom built, low temperature microscopy setup\cite{Tian2016RSI}. A 532 nm excitation laser, whose spot has a diameter of 2 $\mu m$, was used with the power limited to 10 $\mu$W to minimize sample heating while allowing for a strong enough signal. At both room and base temperature (10K), the reported spectra were averaged from three spectra in the same environment to ensure reproductability. The spectrometer had a 2400 g/mm grating, with an Andor CCD, providing a resolution of $\approx 1$ cm$^{-1}$. Dark counts are removed by subtracting data collected with the same integration time with the laser blocked. 
Freshly cleaved sample is transferred to a cryostat with optical window and pumped down to a vacuum level of $10^6$ torr. The 488 nm (2.54 eV) emission line from an Argon laser is used as the excitation source. The incident laser is reflected by a 90/10 nonpolarizing cube beam splitter and then focused to a spot size of 2 $\mu$ m on the sample using a 50× objective lens (NA, 0.35). The laser power on the sample is 700uW. The incident beam and  collected optical signal are linear polarization resolved using a combination Fresnel rhomb retarders and linear polarizers. Raman signal is dispersed by a Horiba T64000 spectrometer equipped with 1800g/mm gratings and and detected with a liquid-nitrogen-cooled CCD camera. We used fresnel rhomb to measure the angular dependent Raman spectra in both setup.

\section*{Acknowledgement}
We thank L. Benfatto and A. Chubukov for useful discussions about the CDW Raman response. \textbf{Funding:} Y.W. is grateful for the support of the for the support of the Office of Naval Research under Award number N00014-20-1-2308. The work of M.M.H. was supported by the Air Force office of Scientific Research under award number FA9550-20-1-0246. The work of G.M. was supported by the National Science foundation via award DMR-2003343. K.S.B. acknowledges the support of the the U.S. Department of Energy (DOE), Office of Science, Office of Basic Energy Sciences under award no. DE-SC0018675. L.M.S. acknowledges support from the Gordon and Betty Moore Foundation through Grant GBMF9064, the David and Lucile Packard foundation and the Sloan foundation. J.J.C and J.L.H. gratefully acknowledge the support from the Gordon and Betty Moore Foundation (EPiQS Synthesis Award). Y.-C. W. and J. Y. are supported by National Science Foundation under award number DMR-2004474. Work by D.X is supported by DOE Award No. DE-SC0012509. Work by I.P. and P.N. was partially supported by the Quantum Science Center (QSC), a National Quantum Information Science Research Center of the U.S. Department of Energy (DOE). I.P. is supported by the Swiss National Science Foundation (SNSF) under project ID P2EZP2\_199848.  P.N. is a Moore Inventor Fellow and gratefully acknowledges support through Grant GBMF8048 from the Gordon and Betty Moore Foundation.
\textbf{Author contributions:} Y.W. performed the Raman experiments and analyzed the data. G.M helped with data fitting and plotting. S.L, L.S. grew the \rte\ crystals. Y-C.W and J.Y helped with 488nm Raman measurement. H.L and D.X performed the DFT calculations. J.L.H. and J. J. C. performed TEM measurement. I.P. and P.N. developed the theory. Y.W and M.H wrote the manuscript with the help of K.S.B. K.S.B.conceived and supervised the project. \textbf{Competing interests:} The authors declare no competing interests. \textbf{Data and materials availability:} All data are available in the main text or the supplementary materials.The data that support the plots within this paper and other findings of this study are available from the corresponding authors on request. 

\section*{Data Availability Statement}
The datasets generated during and/or analysed during the current study are available from the corresponding author on reasonable request.

\begin{figure}[h]
\includegraphics[scale= 0.73]{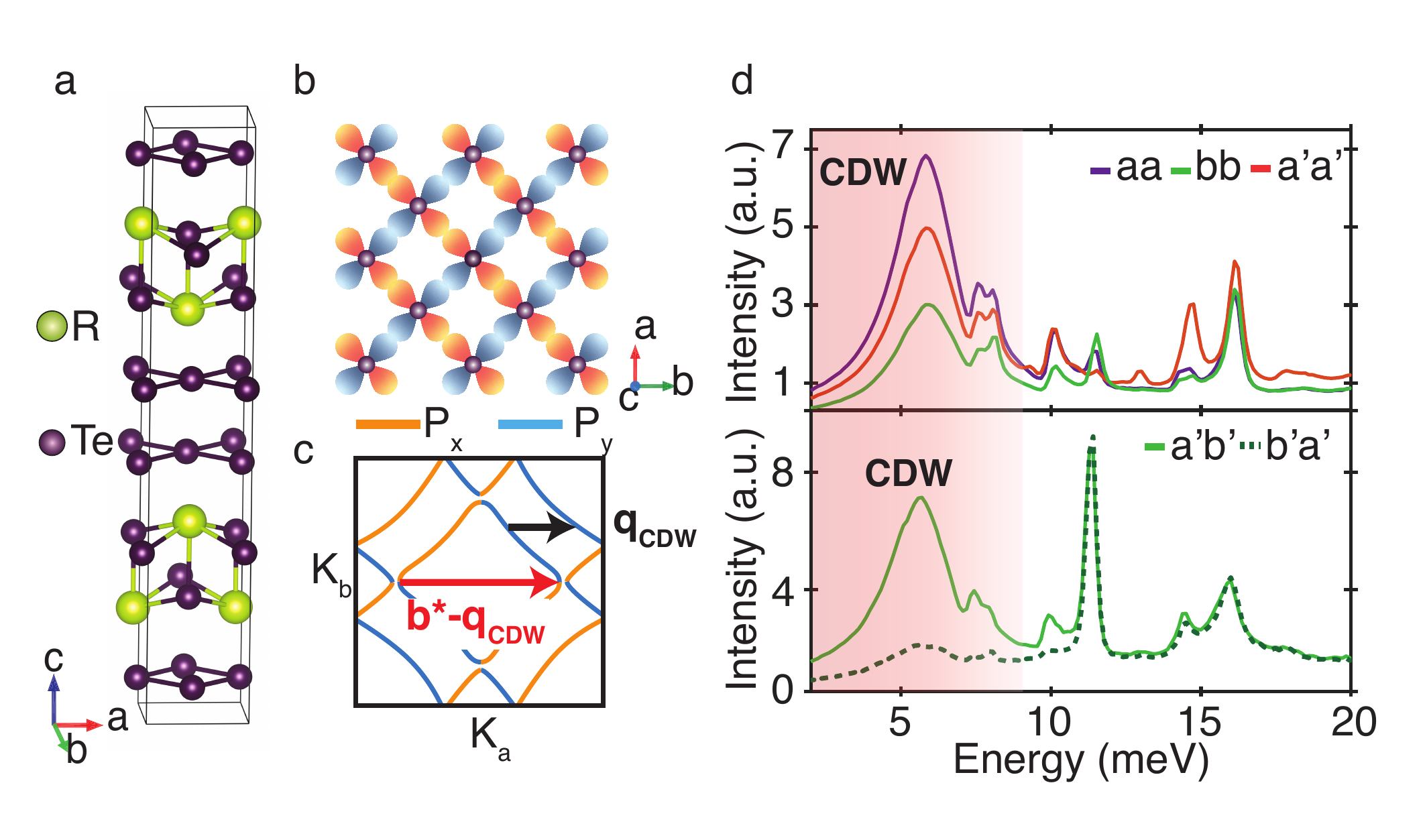}
\caption{\textbf{\rte\ structure and representative Raman spectra} (a) \rte\ crystal structure and unit cell (black line). (b) $P_{X}$ (orange) and $P_{y}$ (blue) orbitals in the Te layer. (c) Fermi surface with orbital content labeled by the same colors as in (b). The red arrow indicates the CDW vector ($q_{CDW}$) with the black arrow indicating the second nesting condition with the reciprocal lattice vector ($b^{*}-q_{CDW}$) (d)300K Raman spectra of \gdte. Top plot is taken in parallel linear polarization, with incident and scattered light aligned with different crystal axis. The Higgs mode is shaded. The bottom plot is taken in cross linear polarization, for incident light aligned with the a'(45 deg off a-axis) direction and scattered light along b'(45 deg off a-axis) direction (green solid line). Upon swapping the incident and scattered polarization (dashed line), the response of all phonons modes is identical, while the amplitude mode is suppressed. }
\label{fig:1}
\end{figure}

\begin{figure}[h]
\includegraphics[scale=0.8]{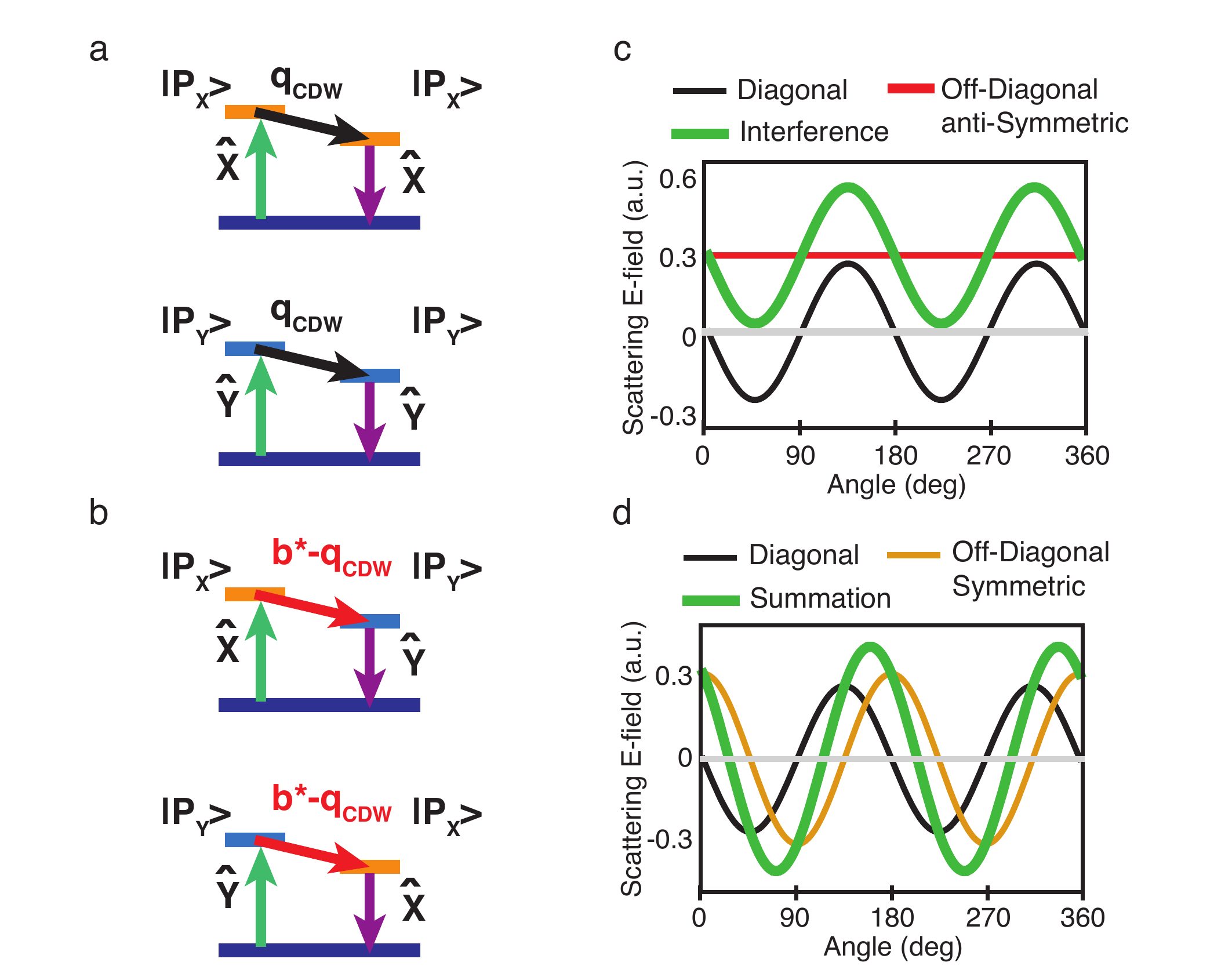}
\caption{\textbf{Interference of quantum pathways} (a) CDW involved symmetric Raman scattering process. (b) CDW involved anti-symmetric Raman scattering process. (c) Scattered electric field resulting from quantum interference of  a diagonal (R$_{ii}$) and antisymmetric off-diagonal ($R_{ij}=-R{ji}$) Raman processes. (d) Scattered electric field resulting from quantum interference of  a diagonal (R$_{ii}$) and symmetric off-diagonal ($R_{ij}=R{ji}$) Raman processes. The presented angle is the polarization rotation angle relative to the crystal a-axis.}
\label{fig:2}
\end{figure}

\begin{figure}[h]
\includegraphics[scale=0.72]{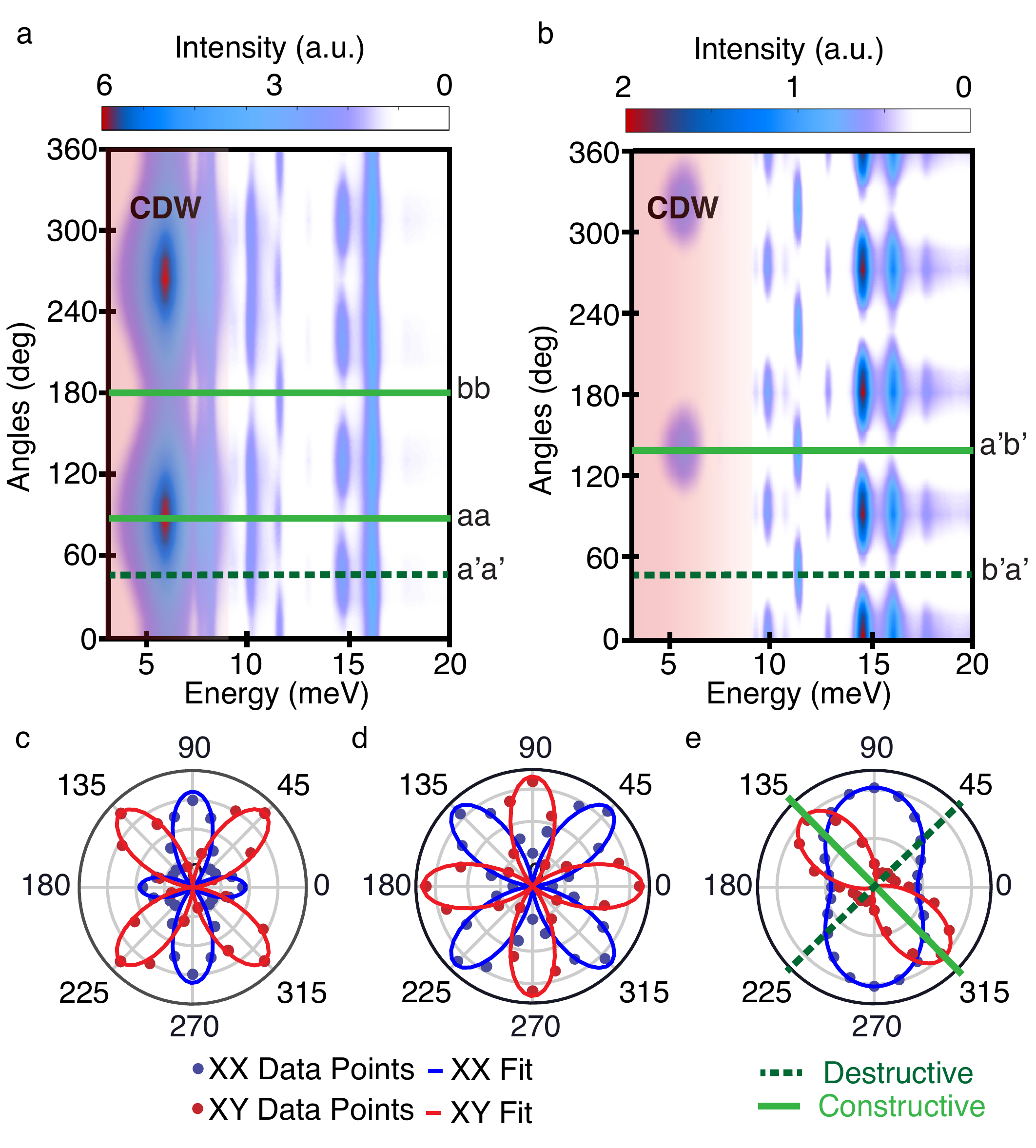}
\caption{\textbf{Angular resolved Raman intensities} Color map of angular resolved Raman at 300K in parallel (a) and cross (b) linear polarization. The green lines indicate the angles of the representative spectra in Fig.1D. (C-E)Angular dependence of the ampltiudes of the Raman modes extracted from Voigt fits of the spectra in parallel (red dots) and cross (blue dots) linear polarization. (c) A$_{g}$ mode of \gdte.  (d) B$_{g}$ mode of \gdte. (e) CDW mode of \gdte\, revealing the constructive (green a'b') versus destructive (dashed green b'a') interference.}
\label{fig:3}
\end{figure}

\begin{figure}[h]
\includegraphics[scale=0.72]{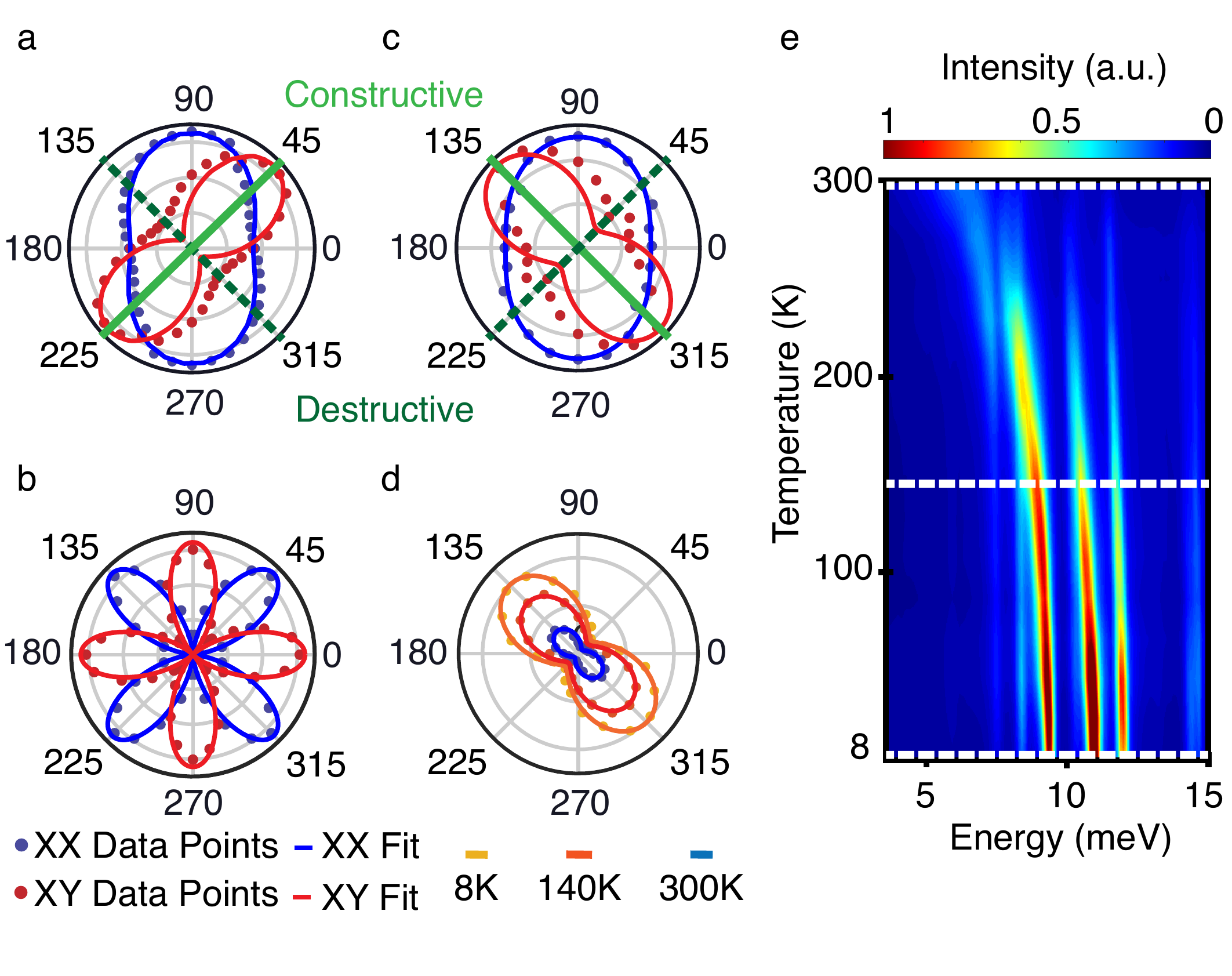}
\caption{\textbf{Additional tests of Quantum Interference in RTe$_{3}$} (a) Angular dependence of the Higgs mode in \late\ measured with a 532 nm excitation laser. (b) $B_g$ mode of \late\ with 532 nm laser following the expected angular dependence. (c) \gdte\ Higgs mode measured with a 488 nm excitation laser on a different system. (d) The angular dependence of the Higgs  intensity (dots) and fitting from Raman tensors(lines) at different temperatures, demonstrating the effect is insensitive to mixing with the neighboring phonon mode. (e) Temperature dependent Raman color map of \gdte.} 
\label{fig:4}
\end{figure}

\end{document}